\documentclass[final, 5p, times, twocolumn]{elsarticle}
\usepackage{amsmath,amssymb,amsfonts}

\usepackage{framed,multirow}

\usepackage{latexsym}

\usepackage{url}
\usepackage{array}

\newcommand{\etal}[0]{\textit{et al.}~}

\newcommand{\Seg}[2]{$\text{S}_\text{#2}^\text{#1}$}
\newcommand{\Trans}[1]{\mathcal{T}_\text{#1}}

\newcommand{\review}[1]{#1}

\makeatletter
\def\ps@pprintTitle{%
  \let\@oddhead\@empty
  \let\@evenhead\@empty
  \let\@oddfoot\@empty
  \let\@evenfoot\@oddfoot
}
\makeatother

\begin{document}
%
\begin{frontmatter}
\title{Measurement and Analysis of Lobar Lung Deformation After a Change of Patient Position During Video-Assisted Thoracoscopic Surgery} %

\author[1]{Pablo Alvarez\corref{cor1}}

\author[1]{Matthieu~Chabanas}
\cortext[cor1]{Corresponding author: matthieu.chabanas@univ-grenoble-alpes.fr
}

\author[1]{St\'ephane Sikora}

\author[2,3]{Simon Rouzé}

\author[1]{Yohan Payan}

\author[2]{Jean-Louis Dillenseger}

\address[1]{Univ. Grenoble Alpes, CNRS, UMR 5525, Grenoble INP, TIMC, 38000 Grenoble, France}
\address[2]{Univ. Rennes 1, Inserm, LTSI, UMR 1099, 35000 Rennes, France}
\address[3]{Department of Cardio-Thoracic and Vascular Surgery, CHU Rennes,35000 Rennes, France}

\begin{abstract}
Video-assisted thoracoscopic surgery (VATS) is a minimally invasive surgical technique for the diagnosis and treatment of early-stage lung cancer. During VATS, large lung deformation occurs as a result of a change of patient position and a pneumothorax (lung deflation), which hinders the intraoperative localization of pulmonary nodules. Modeling lung deformation during VATS for surgical navigation is desirable, but the mechanisms causing such deformation are yet not well-understood. In this study, we estimate, quantify and analyze the lung deformation occurring after a change of patient position during VATS. We used deformable image registration to estimate the lung deformation between a preoperative CT (in supine position) and an intraoperative CBCT (in lateral decubitus position) of six VATS clinical cases. We accounted for sliding motion between lobes and against the thoracic wall 
and obtained consistently low average target registration errors (under 1~mm). We observed large lung displacement (up to 40~mm); considerable sliding motion between lobes and against the thoracic wall (up to 30~mm); and localized volume changes indicating deformation. These findings demonstrate the complexity of the change of patient position phenomenon, which should necessarily be taken into account to model lung deformation for intraoperative guidance during VATS.
\end{abstract}
\end{frontmatter}
%

\section{Introduction}
Lung cancer screening through low-dose CT has significantly increased the detection of pulmonary nodules \cite{dekoning_nelson_2020}. 
To date, the diagnosis of these nodules requires histological analysis of tissue sample \cite{keating_novel_2016}. However, many of these nodules are too small, too deep or not sufficiently dense for transthoracic or bronchoscopic biopsy procedures to be reliable. Consequently, a surgical approach is often preferred, since it allows for complete nodule resection \cite{keating_novel_2016}. In current clinical practice, this is performed through the minimally invasive video-assisted thoracoscopic surgery (VATS), which is one of the main approaches for pulmonary nodule resection \cite{salfity_vats_2020}.

Although the position of pulmonary nodules is well established prior surgery, the lung deformation caused by a pneumothorax (lung deflation) during surgery makes their intraoperative localization very challenging. Moreover, many of these nodules are generally not visible at the lung surface, nor directly palpable through surgical tools \cite{chao_comparison_2018}. 
Consequently, several pulmonary nodule localization strategies are currently used in clinical practice, most of them based on the placement of markers (\emph{e.g.}, hook-wires, micro-coils, dyes) either pre- or intra-operatively \cite{keating_novel_2016}. However, the placement of such markers often requires an adjuvant intervention that is not devoid of complications. The use of intraoperative imaging for pulmonary nodule localization has therefore gained attention \cite{wada_thoracoscopic_2016, rouze_small_2016}. Although the visibility of the nodules may still be unrealiable in some cases (\emph{e.g.}, small nodules with low density), the use of intraoperative imaging opens the door to deformation estimation approaches for surgical navigation, which could greatly benefit the current clinical practice.

Previous works have studied breathing-induced lung deformation for clinical applications such as radiation-therapy \cite{jafari_in-vivo_2021}, ventilation and function assessment \cite{guerrero_2005, jahani_assessment_2014}, or estimation of mechanical properties \cite{hasse_estimation_2018}. Although various authors have put forward deformation estimation strategies for guided surgery in other contexts (\emph{e.g.}, in laparoscopy \cite{heiselman_character_2017} and brain surgery \cite{gerard_brain_2017}), few works have addressed lung deformation estimation for surgical guidance and pulmonary nodule localization during VATS. Such estimation is challenging, since the lung undergoes very large deformation from the preoperative to the surgical setting.
The first cause of deformation is the change of patient position from supine to lateral decubitus, along with the setup of anesthesia and mechanical ventilation. The combination of these factors will be referred to as change of patient position for simplicity. The second cause of deformation is the pneumothorax, which is induced by the surgical incisions.

Uneri \etal introduced a hybrid registration approach combining surface morphing and deformable intensity-based image registration for Cone Beam CT (CBCT) images of inflated and deflated pig lungs \cite{uneri_deformable_2013}. Nakao \etal proposed a surface-based deformable registration approach based on curvature similarity of complete lung surface reconstructions from CT images of beagle-dog lungs \cite{nakao_shape_2019}, and later quantified displacement patterns and strain measurements from the estimated surface deformations \cite{nakao_deformation_2021}. More recently, researchers have also performed clinical studies. Lesage \etal proposed a hyperelastic biomechanical model in which the lung was constrained by external pressure until reaching a desired deflated lung volume \cite{lesage_preliminary_2020}. The model was built from CT acquisitions before and after a pneumothorax in supine position, which resulted from transthoracic needle-biopsies. Maekawa \etal adapted the curvature similarity surface-based approach to a surgical setting using intraoperative CBCT acquisitions of the inflated and deflated lung of patients in lateral decubitus position during open thoracotomy \cite{maekawa_model-based_2020}. Although these works report encouraging results, they account only for the pneumothorax deformation, fully disregarding the change of patient position.

Few studies have accounted for the change of patient position in the estimation of VATS-induced lung deformation. Nakamoto \etal proposed a surgical navigation system in which the change of patient position was approximated via rigid registration and the pneumothorax via surface-based deformable mesh registration \cite{nakamoto_thoracoscopic_2007}. Recently, we proposed a hybrid registration approach mixing deformable image registration and a poroelastic biomechanical model of the lung, which we evaluated in retrospective clinical studies of a single needle-biopsy case \cite{alvarez_biphasic_2019}, and of five VATS cases \cite{alvarez_hybrid_2021}.
The VATS-induced deformation was estimated in two steps: one for the change of patient position and one for the pneumothorax. We obtained lower estimation errors when accounting for the change of patient position and showed that this deformation entails more than simple rigid body motion, which is in accordance with a recent study \cite{little_2021}. Therefore, we believe that a better understanding of lung deformation after a change of patient position can significantly contribute to robust strategies for VATS-induced deformation estimation, and therefore, pulmonary nodule localization. Moreover, this understanding could also benefit the planning of other thoracic surgeries, since a change of patient position is often required \cite{little_2021}.

In this article, we estimate, quantify and analyze lung deformation after a change of patient position during VATS (\emph{i.e.}, after the change of pose, ventilation, and anesthesia). We used a preoperative CT (in supine position) and an intraoperative CBCT of the inflated lung (in lateral decubitus position, before surgical incisions) of six VATS clinical cases, which respectively provide the lung configuration before and after the change of patient position. The deformable registration of these two images provided deformation fields that were analyzed to characterize displacement, sliding motion and deformation patterns of the lung. This article extends our earlier single-case study \cite{alvarez_lung_2018} by using a more robust registration procedure that accounts for local intensity variations and sliding motion, in addition to a more comprehensive characterization of deformation and movement. 

The contributions of this work can be summarized as follows: (i)~we developed a deformable image-based registration method that accounts for sliding motion and local intensity variations for the estimation of lung deformation after a change of patient position during VATS; (ii)~we characterized the estimated deformation fields by means of displacement, sliding and deformation measurements that were extracted per lung lobe; and (iii)~we report and analyze the obtained deformation measurements for six VATS clinical cases.

\section{Materials and methods}
\subsection{Clinical dataset}
This study included six clinical cases of wedge resection via VATS. All patients had a solitary pulmonary nodule detected through CT examination prior surgery. The localization of these nodules was performed during surgery (without marker-placement procedures) under intraoperative CBCT image guidance \cite{rouze_small_2016}. The resections were performed in a hybrid operating room at the Rennes University Hospital, France, under the approval of the local ethics committee (2016-A01353-48 35RC16 9838) and the informed consent of all patients.

\begin{figure}[tb]
	\centering
    \includegraphics[width=.999\columnwidth]{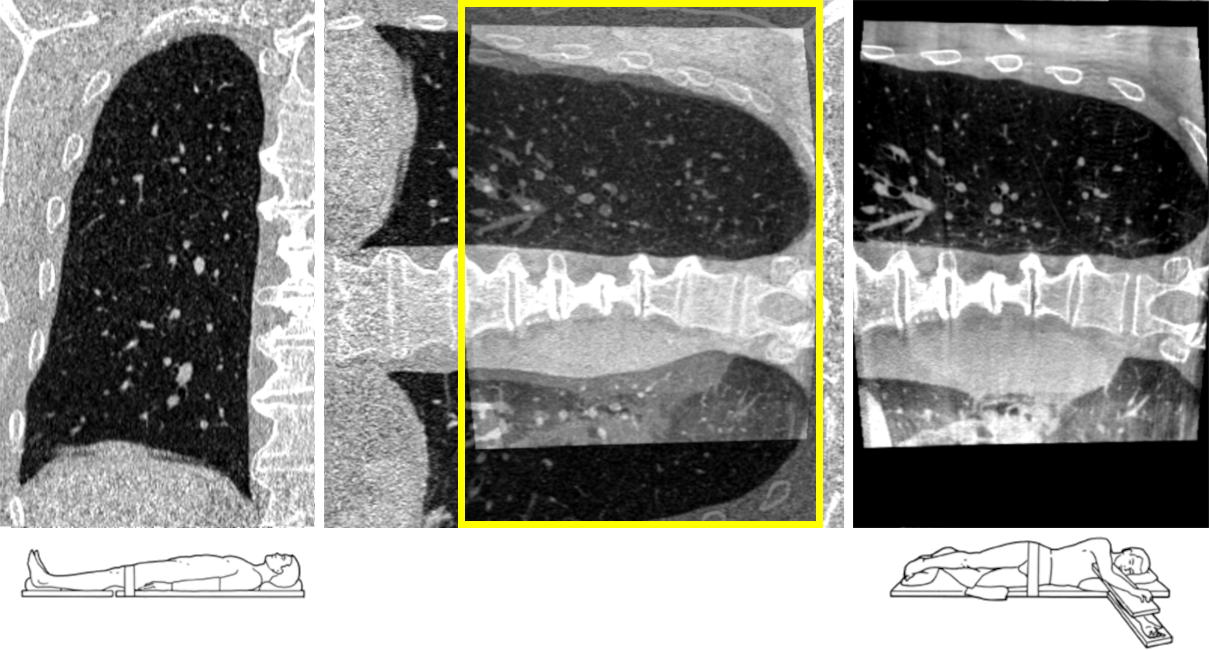}
    \caption{Coronal slices for the preoperative CT and intraoperative CBCT images for Case~1. Left: preoperative CT with the patient in supine position. Right: intraoperative CBCT with the patient in lateral decubitus position. Middle: superposition of the intraoperative CBCT and preoperative CT after rigid registration of the spine. The FOV of the intraoperative CBCT is represented with a yellow rectangle.
    \label{fig:images}}
\end{figure}

Two images were used per patient: a preoperative CT and an intraoperative CBCT. The preoperative CT (0.97~mm of axial plane resolution, 0.8~mm of slice thickness and average size of $512 \times 512 \times 337$ voxels) corresponds to the standard diagnosis image. It was acquired with the patient in supine position and under breath-hold at the end-of-inhalation. The CBCT image (isotropic voxel resolution of 0.48~mm and $512 \times 512 \times 354$ voxels) was acquired during surgery with the patient in lateral decubitus position \review{(operated lung on top)}, under general anesthesia and with mechanical ventilation paused at the end-of-expiration. 
The acquisition was performed with a C-arm system (Artis Zeego, Siemens Healthcare, Germany) before the surgical trocar incisions. The limited field-of-view (FOV) of the C-arm ($\approx$~17~cm in the cranio-caudal axis) provided only a partial view of the operated lung. Coronal slices of these two images are depicted in Fig. \ref{fig:images} for one clinical case.

For validation purposes, paired anatomical landmarks were identified in the two images by a single cardio-thoracic surgeon, in a single sitting per case \cite{alvarez_hybrid_2021}. 
Additional landmarks were added in this study to improve the coverage of structures near the lung fissures. In total, 50 landmarks were available per case \review{(2 to 16 at less than 1~cm from the fissures). An illustration of the landmarks' spatial distribution is provided in the supplementary materials.}

\subsection{Image registration}
Let $\mathcal{F}, \mathcal{M} : \mathbb{R}^3 \to \mathbb{R}$ be the fixed and moving images corresponding to the intraoperative CBCT and preoperative CT, respectively. Let the FOV of $\mathcal{F}$ be denoted by $\Omega_{\mathcal{F}} \in \mathbb{R}^3$. The objective is to find a transformation $\mathcal{T}: \Omega_{\mathcal{F}} \to \mathbb{R}^3$, aligning the fixed $\mathcal{F}$ and moving $\mathcal{M}$ images, such that for all $x\in \Omega_\mathcal{F}$, $\mathcal{F}(x)$ and the deformed $\mathcal{M}(\mathcal{T}(x))$ are similar under a suitable similarity criteria. Note that choosing the intraoperative CBCT as the fixed image $\mathcal{F}$ ensures that all points in $\Omega_\mathcal{F}$ have a correspondence in $\Omega_\mathcal{M}$, since the lung is entirely visible only in the preoperative CT. 

\subsubsection{Transformation model}
The parametric Free-Form Deformation (FFD) model based on B-Spline interpolation \cite{rueckert_nonrigid_1999} was chosen since it has performed well in previous studies \cite{wu_evaluation_2008, kanai_evaluation_2014}. This transformation model is denoted $\mathcal{T}_{\boldsymbol\mu}$. 

\subsubsection{Cost function}
The optimal $\mathcal{T}_{\boldsymbol\mu}$ should maximize the similarity between the fixed $\mathcal{F}(x)$ and deformed moving $\mathcal{M}(\mathcal{T}_{\boldsymbol{\mu}}(x))$ images. To measure such similarity, problem-specific criteria should be considered. For instance, to account for the local intensity changes induced by lung ventilation, authors have proposed locally-invariant metrics based on image intensity gradients \cite{haber_intensity_2006} or patch-based descriptors \cite{heinrich_mind_2012, cachier_3d_2000}. More recently, authors have proposed a mixture of image similarity and boundary similarity \cite{ruhaak_estimation_2017}. 
Explicitly accounting for boundary similarity not only has been shown to improve overall registration \cite{ruhaak_improving_2011}, but also to allow for sliding motion estimation \cite{wu_evaluation_2008, vandemeulebroucke_automated_2012}. Following these previous works, the parameters $\boldsymbol\mu$ of $\mathcal{T}_{\boldsymbol\mu}$ were found by minimizing: 
\begin{equation}
\mathcal{C}(\boldsymbol\mu; \mathcal{F}, \mathcal{M}) := \mathcal{D}(\boldsymbol\mu; \mathcal{F}, \mathcal{M}) + \mathcal{B}(\boldsymbol\mu; \mathcal{F}, \mathcal{M}) + \alpha \mathcal{R}(\boldsymbol\mu), \nonumber
\label{eq:cost}
\end{equation}
where $\mathcal{D}$ is a measure of locally-invariant intensity similarity, $\mathcal{B}$ is a measure of boundary similarity, $\mathcal{R}$ is a regularization term enforcing the smoothness of the transformation and $\alpha$ allows to tune-in the strength of the regularization term. 

The locally-invariant image similarity $\mathcal{D}$ was implemented as a measure of image gradient similarity. However, instead of measuring the alignment of normalized gradient fields as originally proposed by Haber \etal \cite{haber_intensity_2006}, the normalized cross correlation (NCC) of gradient magnitudes was herein used:
\begin{equation}
	\mathcal{D}(\boldsymbol{\mu}; \mathcal{F}, \mathcal{M}) := \mathrm{NCC}(\boldsymbol{\mu} ; \lVert \nabla \mathcal{F} \rVert_{\eta}, \lVert \nabla \mathcal{M} \rVert_{\eta}) \nonumber
\end{equation}
with $\lVert\cdot\rVert_{\eta}$ denoting a magnitude operator allowing the filtering of small image gradients by scaling down magnitudes in the range $[0, \eta]$ with a ramp function of slope $1 / \eta$. \review{Using the NCC allows accounting for offsets in gradient magnitudes between the images, which may occur due to tissue densification.}

The boundary similarity $\mathcal{B}$ typically consists in a measure of overlap between the lung segmentations \cite{ruhaak_improving_2011}. While such strategy has performed well in previous works, we experienced some unrealistic deformation patterns far from the boundary. To alleviate this issue, the boundary alignment similarity $\mathcal{B}$ was then defined as:
\begin{equation}
\mathcal{B}(\boldsymbol{\mu}; \mathcal{F}, \mathcal{M}) := \mathrm{NCC}(\boldsymbol{\mu} ; \mathcal{F}_\mathrm{msk}, \mathcal{M}_\mathrm{msk}) \nonumber
\label{eq:boundary_term}
\end{equation}
where $\mathcal{F}_\mathrm{msk}$ and $\mathcal{M}_\mathrm{msk}$ correspond to the fixed and moving images after being masked with their respective segmentations. Following Wu \etal \cite{wu_evaluation_2008}, voxels outside the masks were attributed a constant HU value below the range of possible HU values throughout the lung. 

The masking strategy introduces an artificial intensity gradient at the lung boundary that severely penalizes boundary misalignment during registration. In addition, the image intensities at the interior of the lung may contribute to the measurement of intensity similarity in $\mathcal{D}$. This is especially useful in regions with significant partial-volume effects (\emph{e.g.}, periphery of the lung), where gradient information may be unreliable.

Finally, a physically plausible transformation of a continuous structure needs to be smooth and without foldings. In this study, the bending energy regularizer was used to penalize strong changes in local transformation curvature, as:
\begin{equation}
\mathcal{R}(\boldsymbol{\mu}) := \frac{1}{\lvert \Omega_\mathcal{F} \rvert} \sum_{\mathbf{x} \in \Omega_\mathcal{F}} \lVert \mathrm{\mathbf{H}}_{\mathcal{T}_{\boldsymbol{\mu}}} \rVert_\mathrm{F}^2 \nonumber
\end{equation}
where $\mathrm{\mathbf{H}}_{\mathcal{T}_{\boldsymbol{\mu}}} \in \mathbb{R}^{3\times3\times3}$ is the Hessian matrix of the transformation $\mathcal{T}_{\boldsymbol{\mu}}$ and $\lVert \cdot \rVert_\mathrm{F}$ denotes the Frobenius norm. The Hessian matrix $\mathrm{\mathbf{H}}_{\mathcal{T}_{\boldsymbol{\mu}}}$ contains all the second partial spatial derivatives of $\mathcal{T}_{\boldsymbol{\mu}}$. Therefore, $\mathcal{R}$ penalizes sharp changes in local transformation curvature.
\begin{figure*}[tb]
	\centering
    \includegraphics[width=.9\textwidth]{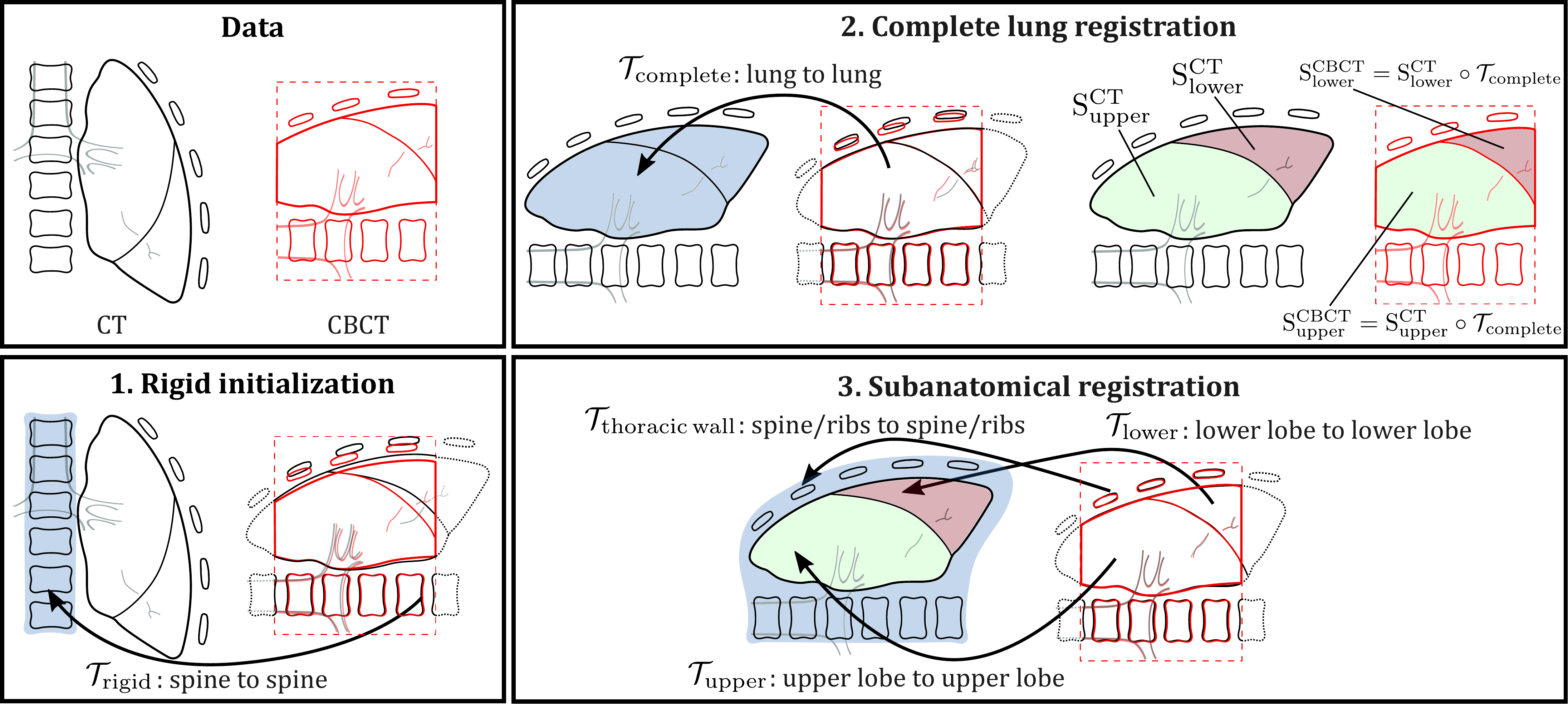}
    \caption{Schematic illustration of the registration process for a left lung. First, rigid initialization over the spine. Second, deformable registration of the complete lung to estimate the lobe segmentations in the CBCT (\Seg{CBCT}{lobes}). Third, deformable registration of the individual lobes and thoracic wall (subanatomical registration). The aggregate transform aligns the thoracic wall and internal lung structures while accounting for sliding motion between the lobes and against the thoracic wall. 
    \label{fig:reg_workflow}}
\end{figure*}

\subsubsection{Implementation}
The registration procedure was implemented into Elastix, an open source toolbox for parametric image registration (\url{https://elastix.lumc.nl}). The cost function $\mathcal{C}$ was minimized with the adaptive stochastic gradient descent method \cite{klein_adaptive_2009}, which estimates $\mathcal{C}$ from random samples within the interesection of $\Omega_\mathcal{F}$ and the deformed $\Omega_\mathcal{M}$. A total of 5000 samples were used in this work. 

A multi-resolution registration strategy was adopted to cope with large deformation while avoiding local minima. Five resolutions were used with image down-sampling factors of $(1/2)^n$, $n$ varying from 4 to 0. The same multi-resolution schedule was applied to the FFD grid, with a minimum size of 16~mm. The FFD grid size was chosen experimentally, not too large to allow sufficient degrees-of-freedom, but not too small to geometrically limit transformation folding \cite{choi_injectivity_2000}.

\subsection{Image segmentation}
\label{sec:segmentations}
Anatomical segmentations of the lung, lobes and spine were needed to define regions-of-interest (ROI) for the computation of each individual term of the registration cost function $\mathcal{C}$. While fully-automatic CT segmentation methods have been proposed to segment the lung fissures \cite{gerard_2019} or lobes \cite{park_lobes_2020}, it is unclear whether these methods would be adequate to segment reduced-quality CBCT images with partial lung visibility. Therefore, we used a semi-automatic process, since we considered more important to ensure the correctness of the segmentations than to rely on fully-automatic methods.

The segmentations of the lung and spine were performed as in \cite{alvarez_hybrid_2021}. In the preoperative CT, the lung was segmented with an in-house modified version of Chest Imaging Platform (\url{www.chestimagingplatform.org}) that is based on automatic region growing and thresholding. This segmentation was then deformed into the CBCT via intensity-based image registration and manually corrected when necessary. The spine segmentation was needed only for the CBCT. It was computed using thresholding over a semi-automatically detected ROI surrounding the spine. In the following, these three segmentations will be referred to as \Seg{CT}{lung}, \Seg{CBCT}{lung} and \Seg{CBCT}{spine}, respectively.

The segmentation of individual lobes was performed only in the CT. First, the lung fissures were approximated with a thin-plate spline fitted over a set of manually placed points (around 100 per fissure). The approximated surfaces served to subdivide the lung segmentation \Seg{CT}{lung} into lobe segmentations, which were subsequently manually corrected when necessary. These lobe segmentations will be referred to as \Seg{CT}{upper}, \Seg{CT}{lower} and \Seg{CT}{middle}, respectively, or jointly as \Seg{CT}{lobes}.

\subsection{Deformation estimation procedure}
We developed a three-step registration procedure to estimate the deformation associated to the change of patient position from the preoperative CT to the intraoperative CBCT: (i)~rigid initialization over the spine, (ii)~deformable registration of the complete lung, and (iii)~deformable registration of the individual lobes and thoracic wall (henceforth referred to as subanatomical registration). 
This registration procedure is illustrated in Fig.~\ref{fig:reg_workflow}. 

\subsubsection{Rigid initialization}
\label{sec:reg_initialization}
The initial alignment of the  images before registration is of little importance as long as the desired registration accuracy is reached.
However, in this study, it could not be chosen arbitrarily as the main goal was not to accurately register the images but to characterize the deformation associated to the change of patient position. Moreover, an anatomical reference was also needed to facilitate comparisons across cases. Since the deformation of the spine observed after a change of patient position was small (see Fig.~\ref{fig:images}), rigid registration of the spine was used for initialization.

To avoid a mismatch among spinal disks, the preoperative CT was first manually aligned to the intraoperative CBCT by applying an axial 90° rotation and a translation to roughly overlap two corresponding spinal disks. Then, rigid registration was performed. This required only the intensity term $\mathcal{D}$, computed inside \Seg{CBCT}{spine} with $\eta=400$ to take advantage of the large gradients between the bone and soft tissue. 

\subsubsection{Complete lung registration}
\label{sec:reg_complete}
After rigid initialization, deformable registration of the complete lung was performed. The intensity term $\mathcal{D}$ was computed inside \Seg{CBCT}{lung}, using $\eta=50$ to remove noisy gradient information. The boundary $\mathcal{B}$ and the regularization $\mathcal{R}$ terms were computed inside \Seg{CBCT}{lung} \review{morphologically dilated} by 5~mm to include boundary information. 
This step allowed estimating the lobe segmentations in the intraoperative CBCT \Seg{CBCT}{lobes}. This was achieved by warping \Seg{CT}{lobes} with the resulting transformation $\Trans{complete}$.

\subsubsection{Subanatomical registration}
\label{sec:reg_subanatomical}
By definition, $\mathcal{T}_{\boldsymbol\mu}$ is only capable of modeling continuous deformation. However, sliding motion may occur between the lobes and against the thoracic wall \cite{schmidt-richberg_estimation_2012, amelon_measure_2014}. Following previous works \cite{wu_evaluation_2008, vandemeulebroucke_automated_2012}, sliding motion was taken into account by independently registering the lobes and the thoracic wall. This required the segmentation of each structure, which was performed manually for the CT and automatically, through registration, for the CBCT. Provided that the accuracy of registration was verified, the quality of the CBCT segmentations was considered sufficient.

The registration of each individual lobe was performed following the procedure described in Sec. \ref{sec:reg_complete}, but using the corresponding lobe segmentations \Seg{CBCT}{lobes}. For the registration of the thoracic wall, the intensity similarity term $\mathcal{D}$ was computed inside \Seg{CBCT}{spine} and \Seg{CBCT}{lung}, both \review{morphologically dilated} by 10~mm to include the ribs, and setting $\eta=400$ to exploit the large bone-tissue gradients\review{, in particular around the ribs. Note that the ribs were never explicitly segmented in any image.} The boundary similarity $\mathcal{B}$ was
computed as the sum-of-squared differences between \Seg{CBCT}{lung} and the deformed \Seg{CT}{lung} instead of the NCC over the masked images, to avoid including unwanted lung intensity information. 

This subanatomical registration step resulted in several elastic transformations, each one corresponding to a different anatomical structure: $\Trans{thoracic\,wall}$, $\Trans{upper}$, $\Trans{lower}$ and $\Trans{middle}$, respectively, the latter only for right lungs.

\subsection{Characterizing lung deformation}
\subsubsection{Surface reconstructions}
\label{sec:surface_reconstruction}
To facilitate the visualization of lung deformation and underlying measurements, 3D surface reconstructions of the undeformed lung geometry were generated. 
These reconstructions were computed from the anatomical segmentations \Seg{CT}{lobes} after rigid initialization. \review{The resulting surfaces have regularly sized triangles of approximately 3~mm.}

\begin{figure}[tb]
	\centering
	\includegraphics[width=.815\columnwidth]{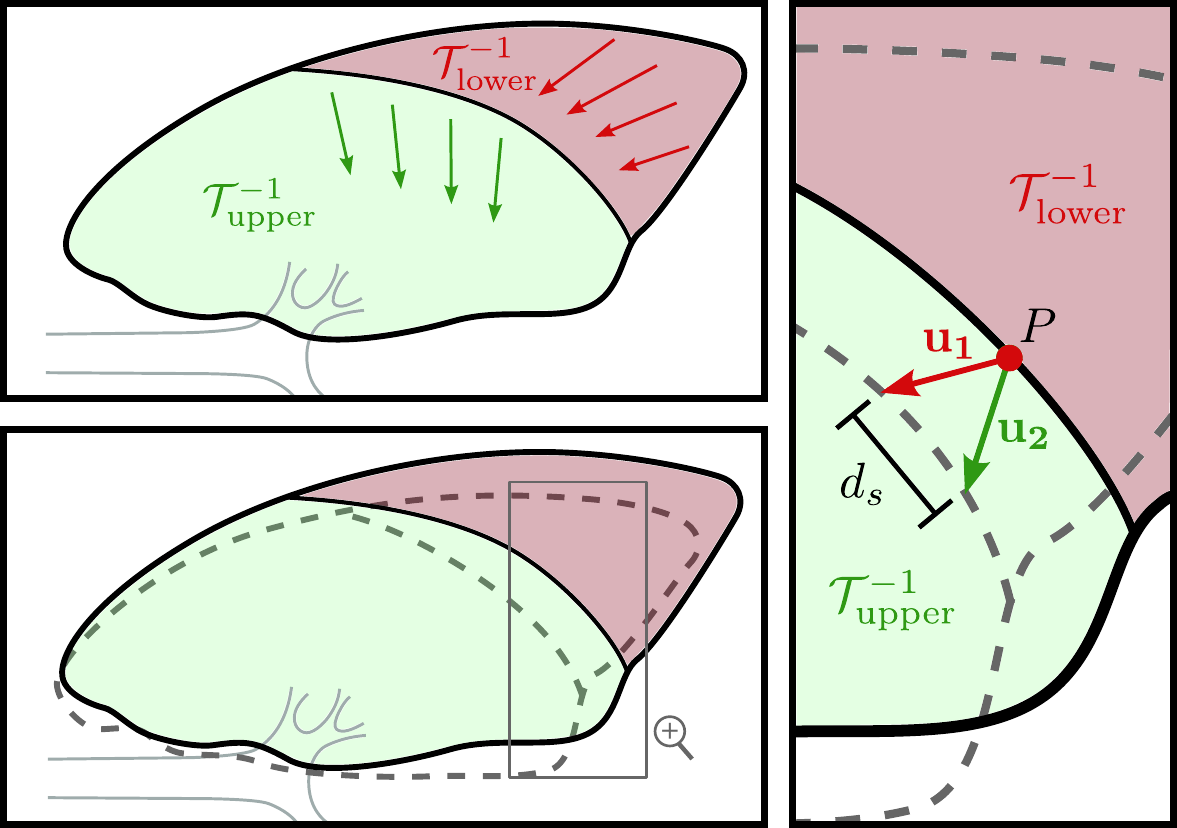}
	\caption{Measurement of sliding motion between two lobes in the undeformed configuration. Each lobe has an associated transformation, $\Trans{upper}^{-1}$ and $\Trans{lower}^{-1}$, respectively. A point lying between the two lobes has two associated displacement vectors $\mathbf{u_1}$ and $\mathbf{u_2}$, each one corresponding to one lobe's transformation. The sliding motion $d_s$ is estimated as the magnitude of the difference between these two vectors. 
	\label{fig:sliding_motion}}
\end{figure}

\subsubsection{Displacements and principal directions}
For visualization purposes, displacements were computed at points inside the undeformed lung geometry. 
These displacements correspond to the inverse displacements estimated by the subanatomical registration process, and were computed via the point-wise iterative algorithm proposed by Crum \etal \cite{crum_methods_2007}.

To study the principal directions of displacement, principal component analysis (PCA) was performed over the inverse deformation fields of each independent lobe. The procedure used followed the work of Hartkens \etal \cite{hartkens_measurement_2003}, who calculated the main directions of a displacement field corresponding to deformation after brain surgery. First, each displacement was represented as a point in 3D-dimensional space (\emph{i.e.}, only the magnitude and direction of displacements are analyzed, not their original positions). The principal directions of displacement were estimated as the directions of maximum variance of the resulting point cloud. In contrast to the traditional PCA algorithm, the variance was estimated from the origin of the space and not from the center of the point cloud, since the latter approach would modify the magnitudes and directions of displacements. The estimated principal directions were scaled with their associated variances for visualization purposes. 

\subsubsection{Jacobian of deformation}
From a continuum mechanics standpoint, the deformation between a reference configuration and deformed configuration can be completely described by the deformation gradient tensor $\mathbf{F}$
\begin{equation}
	\mathbf{F} := \nabla_{\mathbf{X}} \mathcal{T}, \nonumber
	\label{eq:deformation_gradient}
\end{equation}
where $\nabla_{\mathbf{X}}(\cdot)$ represents the gradient operator with respect to the reference coordinates $\mathbf{X}$ and $\mathcal{T}$ a mapping from the reference to the deformed configuration. Here, the reference and deformed configurations correspond to the lung before and after change of patient position, respectively. 

The Jacobian of deformation $J$, defined as $J := \mathrm{det}(\mathbf{F})$, encodes information about local volume changes after deformation. Values of $J$ represent volume contraction ($J < 1$), volume conservation ($J = 1$) and volume expansion ($J > 1$).

We recall that the subanatomical registration procedure estimates a set of mappings $\mathcal{T}_{\boldsymbol{\mu}}$ from the deformed ($\Omega_\mathcal{F}$) to the reference ($\Omega_\mathcal{M}$) configurations. Therefore, the associated deformation gradient tensor corresponds to $\mathbf{F}^{-1}$ and the Jacobian of deformation was computed as:
\begin{equation}
J := 1 / \mathrm{det}(\mathbf{F}^{-1})   \nonumber
\label{eq:jacobian}
\end{equation}

\subsubsection{Sliding motion}
The estimated displacement field may be discontinuous at the interfaces across subanatomical structures, where sliding may occur. 
A point $P$ lying on an interface (\emph{e.g.}, between the upper and lower lobes) in the undeformed configuration has two associated displacements $\mathbf{u_1}$ and $\mathbf{u_2}$, which correspond to the deformations of the underlying anatomical structures sharing the interface. As illustrated in Fig.~\ref{fig:sliding_motion}, the sliding motion $d_s$ was approximated as the magnitude of the difference between these two displacement vectors. This sliding motion was measured at the interfaces across lobes and at the interfaces between the lobes and the thoracic wall. 

\section{Results}
Figures \ref{fig:warping_displacement} to \ref{fig:sliding} present qualitative results for two clinical cases. Images for all cases, as well as the distributions of the computed measures, are available in the supplementary materials.

\begin{figure}[tb]
	\centering
	\includegraphics[width=\columnwidth]{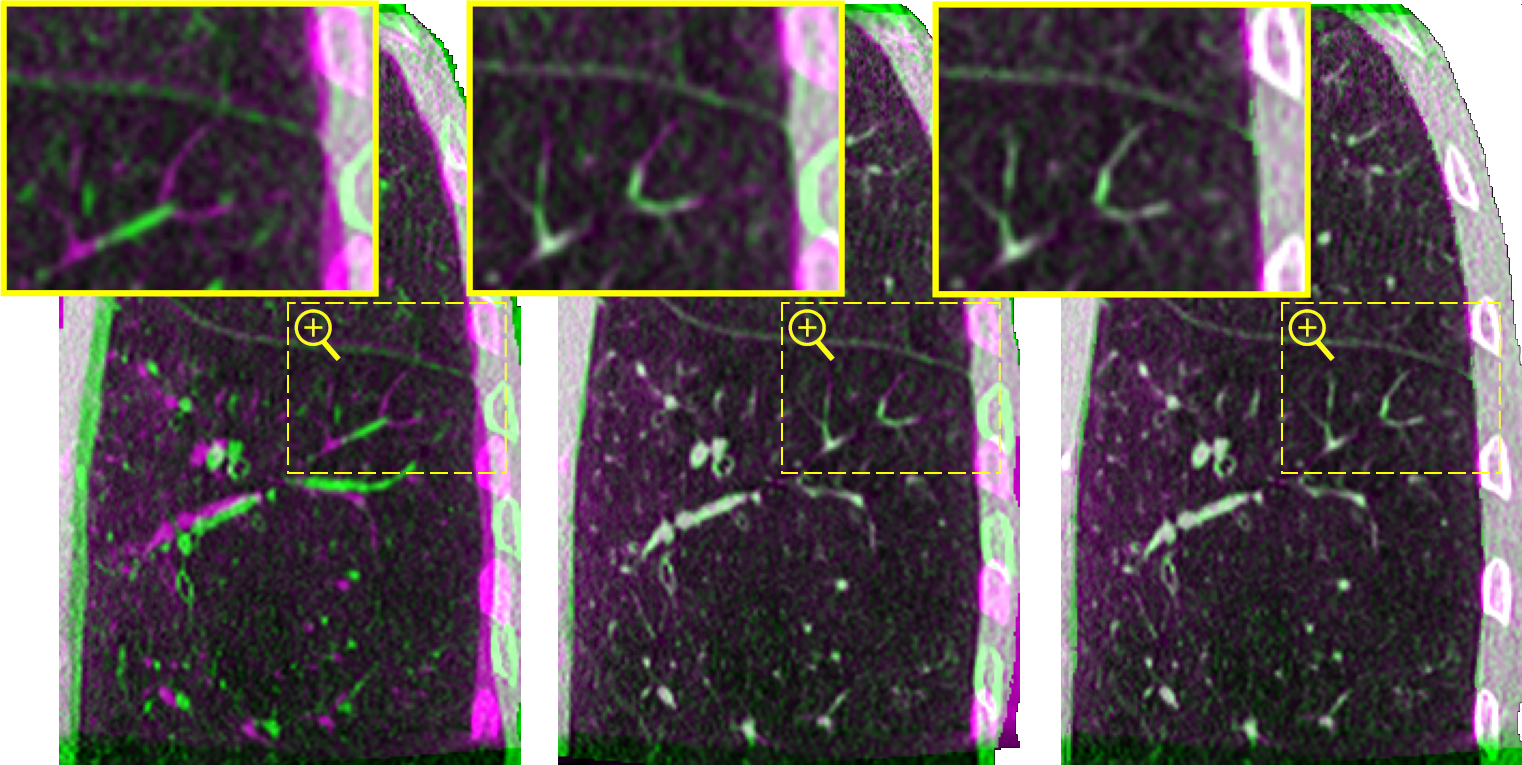}
	\caption{Qualitative registration results for Case 1. Coronal slices showing misalignment after three registration procedures: rigid registration (left), complete lung registration (middle) and subanatomical registration (right). The complementary colors reveal misalignment, with the warped preoperative CT in green and the intraoperative CBCT in magenta. \label{fig:registration_qualitative}}
\end{figure}
\begin{table*}[tb]
	\caption{\label{tab:tre}Quantitative registration results measured with target registration errors (TRE). Mean ($\pm$~standard deviation) TREs are listed for each registration step and each clinical case. Also, TREs are shown jointly for all clinical cases for two groups of landmarks: those far ($\ge 1$~cm) and those close ($\le 1$~cm) to a lung fissure.}
	\centering
	\begin{tabular}{cccccc}
	\hline
	Landmark group & Case & Lung & Rigid initialization & Complete lung registration & Subanatomical registration \\
	\hline
	\multirow{6}{*}{All landmarks} & 1 & Left & 6.7~mm ($\pm$ 3.0~mm) & 1.2~mm ($\pm$ 0.9~mm) & 0.9~mm ($\pm$ 0.5~mm) \\
		& 2 & Right & 12.3~mm ($\pm$ 4.1~mm) & 1.1~mm ($\pm$ 0.7~mm) & 0.8~mm ($\pm$ 0.4~mm) \\
		& 3 & Right & 14.0~mm ($\pm$ 3.2~mm) & 1.1~mm ($\pm$ 0.9~mm) & 0.9~mm ($\pm$ 0.4~mm) \\
		& 4 & Right & 26.6~mm ($\pm$ 4.8~mm) & 1.0~mm ($\pm$ 1.2~mm) & 1.0~mm ($\pm$ 1.1~mm) \\
		& 5 & Left & 18.0~mm ($\pm$ 7.2~mm) & 1.1~mm ($\pm$ 1.0~mm) & 0.8~mm ($\pm$ 0.4~mm) \\
		& 6 & Right & 16.0~mm ($\pm$ 4.8~mm) & 1.0~mm ($\pm$ 0.5~mm) & 1.0~mm ($\pm$ 0.4~mm) \\
	\hline
	\hline
	\textbf{Far} from fissures & All & - & 6.4~mm ($\pm$ 3.2~mm) & 0.9~mm ($\pm$~0.7~mm) & 0.9~mm ($\pm$~0.6~mm) \\
	\hline
	\textbf{Close} to fissures & All & - & 7.3~mm ($\pm$ 2.7~mm) & \textbf{1.5~mm ($\pm$~1.2~mm)}$^{\,\dag}$ & \textbf{0.9~mm ($\pm$~0.5~mm)}$^{\,\dag}$ \\
	\hline
	\vspace{-7pt}\\
	\multicolumn{6}{@{}l}{$^{\,\dag}\,$Statistically significant difference ($p = .001$, non-parametric Wilcoxon sign-ranked test)} \\
	\end{tabular}
\end{table*}

\subsection{Registration accuracy}
The quality of registration was assessed after each one of the three registration steps. Figure \ref{fig:registration_qualitative} depicts coronal slices of the resulting image alignment for a clinical case.
Qualitatively, it is clear that rigid registration was not sufficient to account for the complex lung deformation after a change of patient position. The use of deformable complete lung registration corrected most misalignment, but failed to account for the sliding motion. This is evident from the shifting of the ribs in the thoracic wall, as well as the small structures near the fissure. These issues got corrected with the deformable subanatomical registration while maintaining overall registration accuracy, including the alignment of the lobe boundaries.

Table~\ref{tab:tre} lists target registration errors (TRE) between the paired anatomical landmarks after each registration step. Regardless of the misalignment after rigid initialization, the TREs were consistently reduced to an average of 1.2~mm or less after deformable registration. To demonstrate the importance of accounting for sliding motion across lobes, the landmarks were separated into two groups: one for those closer than 10~mm to a fissure and one for those farther. The subanatomical registration step resulted in a statistically significant improvement of TREs for the landmarks located close to a fissure.

\subsection{Analysis of the deformation}
\begin{figure}[tb]
	\centering
	\includegraphics[width=\columnwidth]{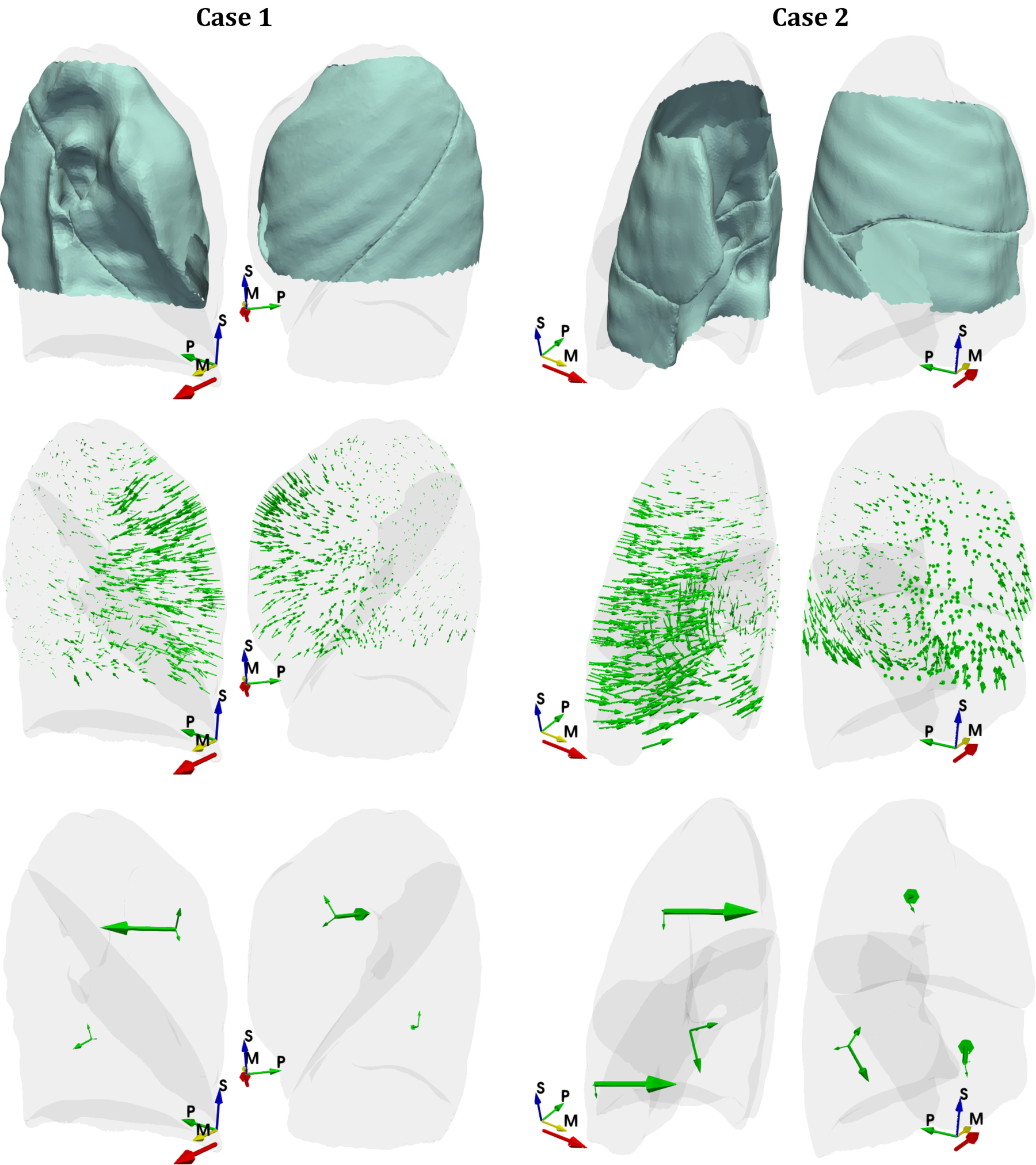}
	\caption{Estimated lung deformation after the change of patient position. Top row, undeformed (translucent) and deformed (plain) surface reconstructions of the lung. Middle row, displacement field of the underlying lung deformation. Bottom row, principal displacement directions per lobe scaled with their relative significance (green arrows). The axes of the CBCT \review{are illustrated with arrows indicating the superior (S), posterior (P) and medial (M) anatomical directions. The direction of gravity after the change of patient position is also indicated by an arrow bellow the axes}. The illustrations for the remaining cases can be found in the supplementary materials. \label{fig:warping_displacement}}
\end{figure}

Figure~\ref{fig:warping_displacement} illustrates the estimated lung deformation after the change of position, as well as the associated displacements and principal displacement directions. The undeformed and deformed lung configurations (top row of Fig.~\ref{fig:warping_displacement}) reveal a contraction motion pattern towards the mediastinum and the spine, \review{as well as the lack of overlapping/penetration across lobar interfaces after deformation confirming the good alignment of lobar fissures.} The corresponding displacement field exhibits more complex motion patterns with significant local variations, both in amplitude and direction of displacements. In general, the amount of displacement appears more important in the anterior and basal regions of the lung, \emph{i.e.}, near the sternum and the diaphragm. Moreover, important differences in displacement direction appear near some lung fissures, for instance, for Case~2, where the right lower lobe moves downwards while the right middle and upper lobes move towards the mediastinum. The principal directions of displacement further reveal these differences across lobes, both in displacement amplitude and direction.

\begin{figure}[tb]
	\centering
	\includegraphics[width=\columnwidth]{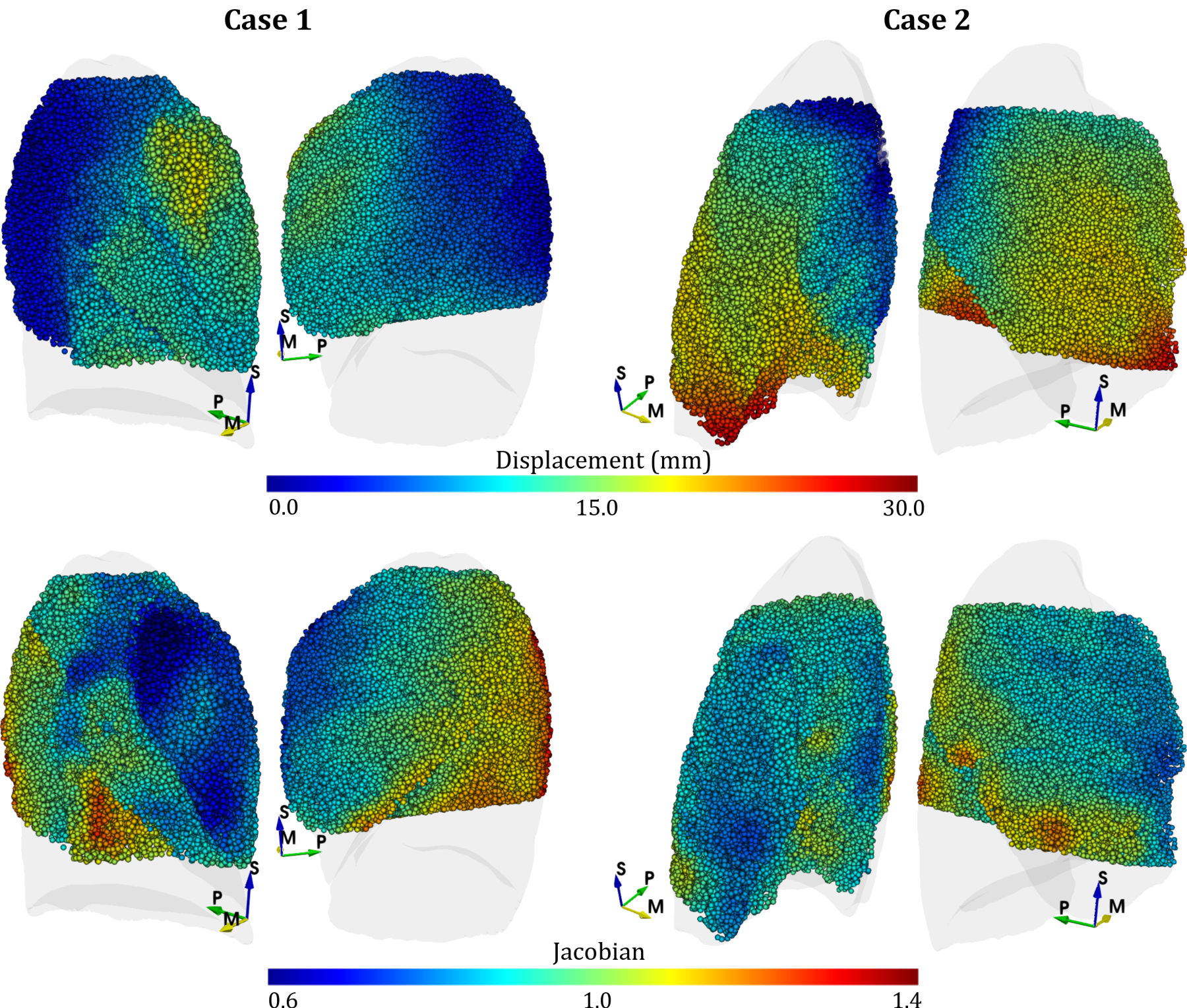}
	\caption{Spatial distribution of magnitude of displacement and jacobian of transformation. The measurements are shown with a color-code at points randomly sampled from the undeformed lung geometry. The axes of the CBCT \review{are illustrated with arrows indicating the superior (S), posterior (P) and medial (M) anatomical directions}. The illustrations for the remaining cases can be found in the supplementary materials. \label{fig:def_measurements}}
\end{figure}

Figure~\ref{fig:def_measurements} illustrates the magnitude of displacement and the jacobian of deformation.
The magnitude of displacement varied significantly across cases with average ($\pm$~standard deviation) values ranging from 7.3~mm ($\pm$~3.6~mm) to 24.6~mm ($\pm$~7.5~mm), and maximal values beyond 40~mm for some cases. Moreover, it varied also across lobes, being significantly larger for the right middle lobe with respect to the other two right lobes ($p < .001$, Welch's unequal variance \textit{t}-test). These differences in magnitude could be observed near the lung fissures, which is an indication of sliding motion across lobes. 

Concerning the jacobian of deformation $J$, generally lower values were located in the extreme anterior region of the lung, near the sternum, which corresponds to the region where the largest displacements were observed. Except for Case~5, the $J$ value was of 0.87 in average across cases, which is an indication of generalized volumetric contraction. Case~5 had a significantly higher average $J$ value of 1.3 ($p < .001$, Welch's unequal variances \textit{t}-test), an indication of generalized volumetric expansion. 
In addition, localized changes of $J$ were present in various regions of the lung, but did not seem to generalize across clinical cases. 

\subsection{Sliding motion}
\begin{figure}[tb]
	\centering
	\includegraphics[width=\columnwidth]{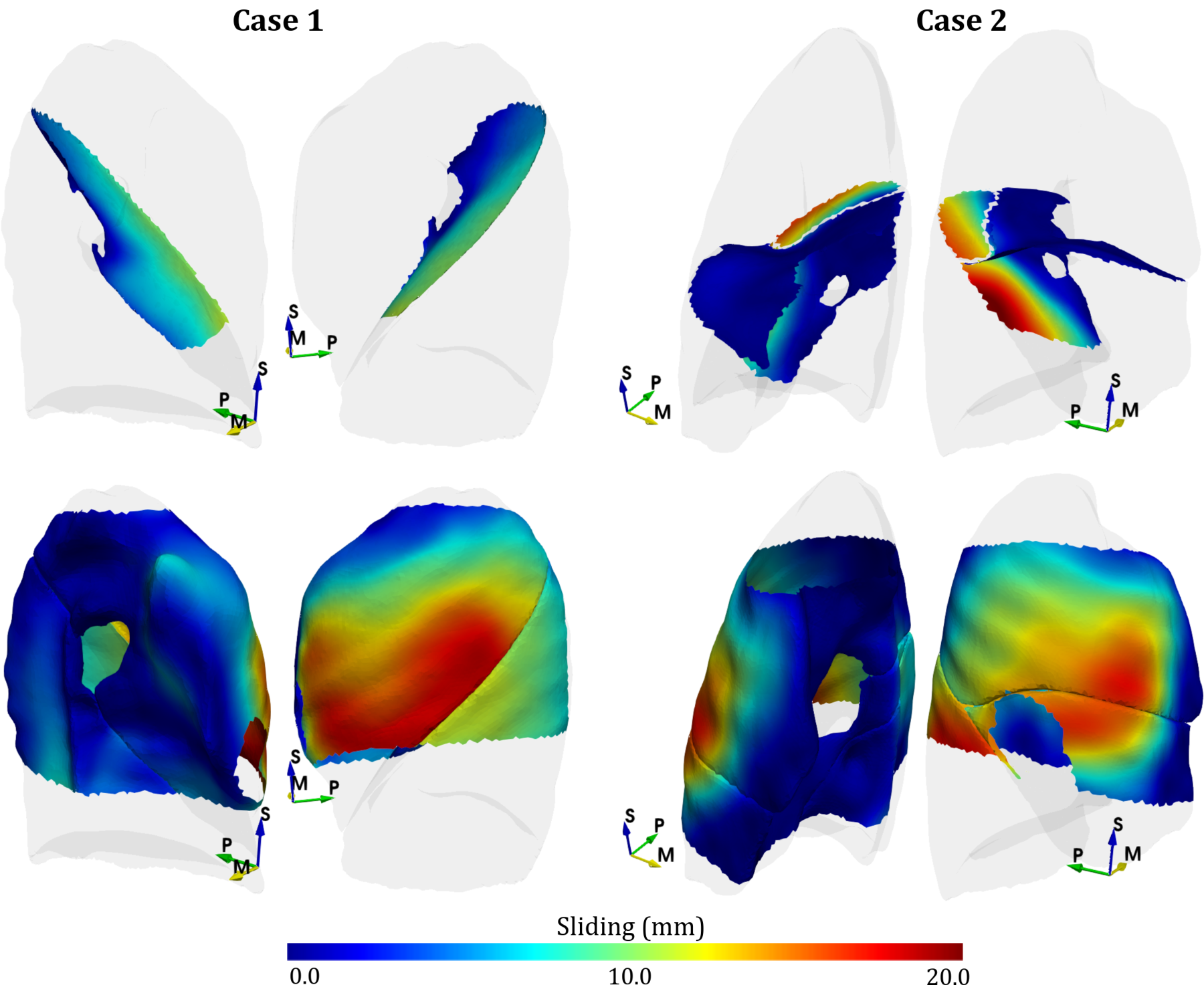}
	\caption{Spatial distribution of sliding motion. First row, sliding motion at the interfaces between lung lobes. Second row, sliding motion at the interface between the lobe surfaces and the thoracic wall. The axes of the CBCT \review{are illustrated with arrows indicating the superior (S), posterior (P) and medial (M) anatomical directions}. The illustrations for the remaining cases can be found in the supplementary materials. \label{fig:sliding}}
\end{figure}

The amount of sliding motion was measured between the lobes and against the thoracic wall. This sliding motion is illustrated in Fig.~\ref{fig:sliding}, where a heat color map is used to represent the magnitude of sliding for each segmented interface. A significant amount of sliding motion was observed across clinical cases, with average ($\pm$~standard deviation) values ranging from 4.1~mm ($\pm$~5.5~mm) to 11.4~mm ($\pm$~7.0~mm) and from 5.9~mm ($\pm$~5.0~mm) to 13.2~mm ($\pm$~8.0~mm), for the sliding motion between lobes and against the thoracic wall, respectively. Except for Case~4, the interface between the right upper and middle lobes presented significantly less sliding motion ($p < .001$, Welch's unequal variances \textit{t}-test). The sliding motion was generally larger in anterior regions of the lung, increasing with the distance to the mediastinum and with the closeness to the diaphragm, with maximal values exceeding 30~mm in some cases. 

\section{Discussion}
\subsection{Registration procedure}
Only accurate deformation fields would allow reliable quantification of deformation and movement measurements. Therefore, the accuracy of the estimated deformation fields was measured in terms of TREs, which were consistently low with maximum average of 1.0~mm across clinical cases. In addition, in comparison with the complete lung registration, the subanatomical registration resulted in significantly lower TREs close to the lung fissures, demonstrating the importance of accounting for sliding motion. The obtained registration accuracy was of the order of the preoperative CT image spacing, which was considered sufficient for the purposes of this deformation characterization study.

\subsection{Displacement and deformation}
The amount of displacement measured was considerable, reaching more than 40~mm for some clinical cases. In general, larger displacements were found in the extreme anterior part of the lung and near the diaphragm. However, the direction of displacement varied significantly across cases, which demonstrates the complexity of the underlying lung deformation. Nevertheless, judging by the principal directions of displacement, we observed considerable displacement towards the mediastinum as well as towards and away from the diaphragm. 
In addition, the right middle lobe presented significantly larger displacement than the other lobes. A plausible explanation could be the position of this lobe with respect to the spine, which was used as the rigid body reference.

Average local volume changes measured through the deformation jacobian $J$ were different from 1.0 across clinical cases. These results suggest that a change of patient position from supine to lateral decubitus during VATS induces overall volume change of the lung. However, it is unclear whether these local volume changes are related to differences in lung ventilation, are a consequence of the deformation induced by the change of patient position, or both. Still, since the mechanical ventilation parameters were approximately the same across clinical cases, the differences observed are likely related to each patient's lung compliance and overall morphology.

During VATS, the diaphragm relaxation following anesthesia \review{and muscle paralysis plays} an important role in lung deformation. Indeed, the relaxed diaphragm moves passively as a result of the forces applied to it (as opposed to actively due to contraction during breathing). Therefore, the diaphragm position gets determined mostly by the interplay between \review{the weight of the abdominal organs and the pressure at the lung inlet imposed by the ventilator \cite{Lohser_2011}, \emph{i.e.}, it is inherently patient-dependent. Since this diaphragm position can significantly determine the resulting lung deformation, it may be a reasonable explanation for the differences observed in lung volume changes, especially for Case~5}. However, this aspect could not be directly studied here because the diaphragm was not visible in the intraoperative CBCT. The influence of diaphragm relaxation in lung deformation after a change of patient position should be investigated in future work.

\subsection{Sliding motion}
A considerable amount of sliding motion was observed, reaching more than 30~mm for some clinical cases. Like for displacement, larger sliding motion appeared towards the anterior and basal regions of the lung, and lower sliding motion appeared towards the mediastinum. Since the lung vessel and airway attachments are located at the mediastinum and lobe fissure incompleteness occurs mostly near the mediastinum \cite{hermanova_incomplete_2014}, the sliding motion is expected to be restricted in this region. In addition, significantly less sliding was observed at the interface between the right upper and right middle lobes, which is a behavior previously reported in the context of lung breathing deformation \cite{amelon_measure_2014}. Interestingly, this interface corresponds to the right horizontal fissure, which is estimated to be incomplete in a significant amount of the adult population (more than 75\% according to He{{\v{r}}}manov{\'{a}} \etal \cite{hermanova_incomplete_2014}). This fissure incompleteness results in a restriction of interlobar sliding, which is a plausible explanation to the reduced sliding motion observed. An exception to this behavior was however observed for Case~4. This particular case had an anatomical variation with the upper and lower lobes completely separated by the middle lobe. It is possible that such difference could result in different deformation patterns, which may explain the increased interlobar sliding observed between the right upper and middle lobes for this case.

It is interesting to note that the clinical cases presenting the largest interlobar sliding were not necessarily those presenting the largest displacements. For instance, Case~4 presented very large displacements 
but relatively low interlobar sliding motion, and the contrary applies for Case~3. This suggests that the interlobar sliding motion is not only deformation-dependent, but also patient-dependent. Indeed, adhesion between lung lobes may develop differently across patients, altering the coefficient of friction at the surface between the lobes, and therefore affecting interlobar sliding motion.

The amount of interlobar sliding motion was found to be as important as the amount of sliding motion against the thoracic wall. This is a crucial insight as many lung motion compensation methods do account for sliding against the thoracic wall but fail to account for interlobar sliding \cite{murphy_evaluation_2011}.

\subsection{Limitations}
An important limitation of this study is the small number of clinical cases, which prevents us to draw statistically significant conclusions. However, the deformation patterns observed were consistent with established surgical knowledge and corroborate clinical observations in other works \cite{little_2021}. Therefore, we believe that our findings are still meaningful and could provide helpful insights towards modeling lung deformation after a change of patient position, which in turn, could foster the development of guidance system for pulmonary nodule localization during VATS.

This study also presents some methodological limitations. First, there is a possible bias due to partially incorrect lobe segmentations. This may be carefully considered as it is well known that the lung fissures are in many cases incomplete \cite{hermanova_incomplete_2014}, and our lobe segmentation algorithm always provided complete fissure estimations. However, fissure incompleteness exists mostly towards the mediastinum, the region where the lungs are attached and therefore less subject to sliding motion. Since the subanatomical registration does not introduce artificial sliding where continuous deformation is expected, the over-segmentation of incomplete fissures should have little effect in the quantification of interlobar sliding motion. Second, most of the lung structures are only partially visible within the FOV of the CBCT image. This implies that all the measurements here presented are not available for the complete structures and therefore should be interpreted carefully, as they are only estimations and may be affected by a sample bias effect. However, the orders of magnitude of displacement, jacobian and sliding motion, as well as their spatial distribution, provide already valuable insights of the undergoing lung deformation after a change of patient position.

\section{Conclusion}
This paper presents a quantification and characterization study of the lung deformation occurring after a change of patient position during VATS. The proposed methodology relied on the deformable registration of a preoperative CT (supine position) to an intraoperative CBCT of the inflated lung (lateral decubitus position). 
The estimated transformations were used to extract displacement, sliding and deformation measurements. To our knowledge, this is the first study to quantify and characterize the lung deformation occurring after a change of patient position during VATS.

Among six clinical cases, we observed large lung displacement reaching 40~mm; considerable sliding motion both between lobes and against the thoracic wall going up to 30~mm; and localized volume changes indicating deformation. Moreover, we observed different directions of displacement across lobes in several cases. Although it is yet unclear whether such lung deformation can be formalized in a deformation model, it is clear that rigid body motion assumptions are insufficient. Starting from a preoperative configuration in supine position, any lung deformation compensation strategy for intraoperative guidance during VATS would require addressing the amount and complexity of the lung deformation induced by a change of patient position. In our future work, we will focus in developing a multi-lobe biomechanical model of lung deformation for intraoperative guidance in thoracic surgery, for which the complex deformation patterns observed in this study will be of great insight.

\section*{Acknowledgements}
This work was supported by the French National Research Agency (ANR) through the frameworks \textit{Images et modèles pour le guidage d’intervention par vidéo-thoracoscopie (VATS) – VATSop} (ANR-20-CE19-0015) and \textit{Investissements d'Avenir Labex CAMI} (ANR-11-LABX-0004).

\bibliographystyle{IEEEtran}
\bibliography{biblio}

\end{document}